\documentclass[%
 reprint,
amsmath,
amssymb,
pra,
floatfix,
balancelastpage,
]{revtex4-1}
\usepackage{color}
\usepackage{epstopdf}
\usepackage{graphicx}
\usepackage{dcolumn}
\usepackage{bm}
\usepackage{amssymb}   
\usepackage{amsmath}
\usepackage{braket}
\usepackage{csquotes}
\usepackage{hyperref}

\begin{document}
\title{Multifaceted nonlinear dynamics in $\mathcal{PT}$-symmetric coupled Li\'{e}nard oscillators}
\author{Jyoti Prasad Deka}\email{jyoti.deka@iitg.ac.in}
\affiliation{Department of Physics, Indian Institute of Technology Guwahati, Guwahati - 781039, Assam, India}
\author{Arjunan Govindarajan}\email[Corresponding author: ]{govin.nld@gmail.com}
\affiliation{Centre for Nonlinear Dynamics, School of Physics, Bharathidasan University, Tiruchirappalli - 620024, India}
\author{Manas Kulkarni}\email{manas.kulkarni@icts.res.in}
\affiliation{International Centre for Theoretical Sciences, Tata Institute of Fundamental Research, Bangalore - 560089, India}
\author{Amarendra K. Sarma}\email{aksarma@iitg.ac.in}
\affiliation{Department of Physics, Indian Institute of Technology Guwahati, Guwahati - 781039, Assam, India}
\begin{abstract}
We propose a generalized parity-time ($\mathcal{PT}$) -symmetric Li\'enard oscillator with two different orders of nonlinear position-dependent dissipation. We study the stability of the stationary states by using the eigenvalues of Jacobian and evaluate the stability threshold thereafter.  In the first order nonlinear damping model, we discover that the temporal evolution of both gain and lossy oscillators attains a complete convergence towards the stable stationary state leading to the emergence of oscillation and amplitude deaths. Also, the system displays a remarkable manifestation of transient chaos in the lossy oscillator while the gain counterpart exhibits blow-up dynamics for certain choice of initial conditions and control parameters. Employing an external driving force on the loss oscillator, we find that the blow-up dynamics can be controlled and a pure aperiodic state  is achievable. On the other hand, the second order nonlinear damping model yields a completely different dynamics on contrary to the first order where the former reveals a conventional quasi-periodic route to chaos upon decreasing the natural frequency of both gain and loss oscillators. An electronic circuit scheme for the experimental realization of the proposed system has also been put forward.
\end{abstract}
\maketitle
\section{\label{sec1}Introduction}
The $\mathcal{PT}$-symmetric formalism in quantum mechanics, originated with the pioneering work of Bender and Boettcher \cite{r1}, has made the former a topic of immense research  interest, not only in quantum mechanics but also in other branches of science and technology \cite{r1a}. It has been demonstrated that non-Hermitian Hamiltonians, which are invariant under the joint operation of the parity and time reversal operators, can exhibit a real eigen spectra. If all the eigenvalues of Hamiltonian are real, it is then said to be in the \textit{unbroken $\mathcal{PT}$-symmetric regime}, else it is said to be working in the \textit{broken $\mathcal{PT}$-symmetric phase}. Under certain parametric changes, all the eigenvalues of such Hamiltonians can coalesce at an exceptional point (EP). This critical point has further been termed as the\textit{ $\mathcal{PT}$-threshold/singularity}.  Subsequently, Ganainy \textit{et al.} \cite{r2} have proposed how optics can facilitate the experimental realization of such non-Hermitian Hamiltonians. It should be noted that such a proposal was possible due to the shared isomorphism between the paraxial equation of diffraction and the time-dependent Schr\"odinger equation. Following these theoretical suggestions, R\"uter \textit{et al.} \cite{r3} have demonstrated the first experimental vindication of $\mathcal{PT}$-symmetry in a configuration of linearly coupled waveguides. Since then, the study of $\mathcal{PT}$-symmetry has been extended in diverse domains and phenomena, which include the onset of chaotic dynamics in optomechanical systems \cite{r4}, modulational instability in complex media \cite{r5}, soliton switch \cite{r6}, active LRC circuits \cite{r7}, optical oligomers \cite{r8,r9,r10,r11,r12,r13,r14,r15}, multilayered structures \cite{r16,r17}, unidirectional reflection-less transmission resonance  in Bragg periodic structures \cite{r18,r19}, wireless power transfer \cite{r20}, plasmonics \cite{r21,r22} and so forth.

In particular, the role of $\mathcal{PT}$-symmetry has been explored in harmonic oscillator models extensively \cite{r7,r33,r33a,r34}. For instance, Carl Bender \textit{et al.} have studied $\mathcal{PT}$-symmetric  phase transition in a system of mechanical oscillators \cite{r33}. Similarly, nonlinearly damped harmonic oscillators were recently considered in the context of $\mathcal{PT}$-symmetry along with other interesting symmetry breaking natures \cite{r33a,r34}. In the present article, we concentrate on the analysis pertaining to a variety of rich nonlinear dynamics of $\mathcal{PT}$-symmetric coupled Li\'enard oscillators. It is important to note that such Li\'enard type oscillators/systems are prototype models to study the dynamics of oscillators with nonlinear dissipation in many areas including physics, electronics and biology \cite{r23}. The nonlinear damping term in Li\'enard systems may cause self-sustained oscillations in the system. Also, these systems comprise of well-known oscillators like Van der Pol oscillator, FitzHugh-Nagumo oscillator, and so on. It is to be noted that the Van der Pol oscillator  gives rise to stable limit cycles based on certain parametric choices and those oscillations are further coined as relaxation-oscillations \cite{r24,r25}. In fact, the Van der Pol oscillator has also been used to model biological regulatory systems such as cardiac and respiratory systems \cite{r26,r27}. Similarly, the FitzHugh-Nagumo oscillator \cite{r28,r29,r30} (which can be reducible to the Van der Pol oscillator) is an another class of Li\'enard systems and it has been used to model nerve impulses. Quite recently, it has been reported that the forced Li\'enard system could be employed to generate extreme events \cite{r31}. On the other hand, recent experimental vindications imply that the nonlinear damping can enhance the quality factor of graphene oscillators significantly \cite{r32}. Moreover, for free falling objects in earth's gravity, the viscous force acting on the object is directly proportional to the square of the velocity and this force leads to the terminal velocity in such objects. Hence, it is obvious that the nonlinear dissipation plays an interesting role in the dynamics of numerous physical  systems.

In this work, we propose a generalized Li\'enard oscillator with balanced (linear and nonlinear) loss and gain. On giving the importance of such oscillators as discussed above, we pay attention on the parity-time symmetric counterpart of such systems. It is well-known that in $\mathcal{PT}$-symmetric coupled oscillators with a linear damping, one oscillator linearly attenuates the perturbation applied to it while the other oscillator amplifies the perturbation by the same proportion \cite{r33}. On the other hand, in a $\mathcal{PT}$-symmetric system with a nonlinear position-dependent dissipation, it must be ensured that in addition to the combination of linear damping and gain, the system should hold the equal amount of nonlinear gain to make it $\mathcal{PT}$-symmetric.  In such nonlinear $\mathcal{PT}$-symmetric systems, we also have considered two cases by taking into account of the higher order version of nonlinear position-dependent dissipation.

The paper is structured  as follows. Section \ref{sec2} deals with the theoretical model of proposed system, which starts with the basics of $\mathcal{PT}$-symmetry in harmonic oscillators following an analytical calculation corresponding to the trace of stationary states of the system. In Sec. \ref{sec3}, we study the stability of the stationary states under variations in gain/loss coefficients. Further, we analyze various nonlinear dynamics of the $\mathcal{PT}$-symmetric Li\'enard system in the neutrally stable region. In Sec. \ref{sec4}, we propose an electronic circuit that can facilitate the experimental realization of our theoretical observations. Finally, we summarize our results along with an outlook in Sec. \ref{sec5}.

\section{MATHEMATICAL MODEL}
\label{sec2}
To start with, let us consider the model of damped harmonic oscillator, which reads as
\begin{equation}
\ddot{x}+\gamma\dot{x}+\omega_0^2x =0
\label{eq1}
\end{equation}
Here, $\gamma$ is the linear damping coefficient and $\omega_0$ is the natural frequency of the oscillator ($\omega_0>0$). Under $\mathcal{PT}$-symmetric operations (i.e. $x\rightarrow-x$ and $t\rightarrow-t$), Eq. \eqref{eq1} transforms to 
\begin{equation}
\ddot{x}-\gamma\dot{x}+\omega_0^2x =0
\label{eq2}
\end{equation}
From Eq. \eqref{eq2}, one can notice that our original damped harmonic oscillator model now becomes an amplified harmonic oscillator model. This implies that a linear harmonic oscillator model will be $\mathcal{PT}$-symmetric only when a damped oscillator is coupled to a gain oscillator.  On the other hand, consider a model of generalized coupled nonlinear harmonic oscillators, which is written as
\begin{subequations}
\begin{align}
& \ddot{x}_1+\gamma\dot{x}_1+\eta x_1^\delta \dot{x}_1+\alpha x_1^3 +\omega_0^2 x_1+\kappa x_2 =0\\
& \ddot{x}_2-\gamma\dot{x}_2-\eta x_2^\delta \dot{x}_2+\alpha x_2^3 +\omega_0^2 x_2+\kappa x_1 =0
\end{align}
\label{eq3}
\end{subequations}
where $\eta$ is the strength of nonlinear damping/gain, $\delta$ is the exponent of nonlinear damping/gain, $\alpha$ is the coefficient of Duffing nonlinearity and $\kappa$ refers to the coupling parameter. Note that in such coupled oscillator systems, the parity ($\mathcal{P}$) and time ($\mathcal{T}$) operators are defined as
\begin{align}
&\mathcal{P}:\quad x_1 \leftrightarrow x_2,\\
&\mathcal{T}:\quad t \rightarrow -t.
\end{align}  
i.e., the two oscillators are swapped under $\mathcal{P}$ operator while the signs of linear and nonlinear gain/loss get reversed under time reversal of $\mathcal{T}$ operator \cite{r33,r33a}. Hence the present novel Li\'enard type systems are  $\mathcal{PT}$-symmetric irrespective of the exponent value of nonlinear damping ($\delta$) as they are invariant upon the combined operation of  $\mathcal{PT}$-symmetry. The pertinent coupled equations can be written in the following way by setting $\dot{x_i}=y_i$
\begin{subequations}
\begin{align}
& \dot{x}_1=y_1 \\
& \dot{y}_1= -\gamma y_1-\eta x_1^\delta y_1- \omega_0^2 x_1- \alpha x_1^3 -\kappa x_2 \\
& \dot{x}_2=y_2 \\
& \dot{y}_2=\gamma y_2+\eta x_2^\delta y_2-\omega_0^2 x_2 - \alpha x_2^3 -\kappa x_1
\end{align}
\label{eq4}
\end{subequations}
In what follows, without loss of generality, we assign all parameters including $\omega_0$ in Eq. \eqref{eq4} to be positive definite. This system admits the following stationary states: 
\begin{enumerate}
  \item The trivial stationary state - $(x_1,y_1,x_2,y_2)=(0,0,0,0)$
 
  \item The non-zero stationary states - $(x_1,y_1,x_2,y_2)=(\pm a_1,0,\mp a_1,0)$, where $a_1=\sqrt {(\kappa-\omega_0^2)/\alpha}$ and $(x_1,y_1,x_2,y_2)=(\pm a_2,0,\pm a_2,0)$, where $a_2=\sqrt {(-\kappa-\omega_0^2)/\alpha}$.
\end{enumerate}
In the next section, we will concentrate on the stability analysis of the above mentioned stationary states and the dynamical aspects of our system for $\delta=1$ and $\delta=2$. To avoid ambiguity, we would like to term our mathematical models as order-1 and order-2 $\mathcal{PT}$-symmetric Li\'enard coupled systems corresponding to the exponent values of nonlinear position dependent gain and loss, respectively. Also, throughout the analysis, we have chosen the following parameter values: coupling constant, $\kappa=0.5$, nonlinear gain/loss, $\eta=0.1$ and Duffing nonlinearity, $\alpha=1$. We will analyze the stability of the stationary states given by, $(x_1,y_1,x_2,y_2)=(a_1,0,-a_1,0)$ (\textbf{FP1}) and $(x_1,y_1,x_2,y_2)=(-a_1,0,a_1,0)$ (\textbf{FP2}), where $a_1=\sqrt {((\kappa-\omega_0^2)/\alpha))}$. It is to be noted that we have not focused on the nontrivial stationary states \textbf{FP3 \& FP4} as they always lead to the saddle node bifurcations without giving rise to any neutrally stable fixed points.

\section{RESULTS AND DISCUSSION}
\label{sec3}
The Jacobian followed by the linearization with respect to an infinitesimal perturbation to the system \eqref{eq4} is given by:
\begin{equation}
J=
\begin{pmatrix}  
0 & 1 & 0 & 0 \\
A & B & -\kappa & 0 \\
  0 & 0 & 0 & 1 \\
-\kappa & 0 & C & D \\
\end{pmatrix}
\label{eq5}
\end{equation}
where $A=-\omega_0^2 -3 \alpha x_1^2 -\delta \eta x_1^{(\delta-1)} y_1$, $B=-\gamma-\eta x_1^\delta$, $C=-\omega_0^2 -3 \alpha x_2^2 +\delta \eta x_2^{(\delta-1)} y_2$ and $D=\gamma+\eta x_2^\delta$. To find the eigen values, the matrix \eqref{eq5} is numerically diagonalized using MATLAB's \emph{eig} package. Also, in the following, in order to have a clear exposition, we discuss various nonlinear dynamics of the system owing to different choice of initial conditions for the two cases of nonlinear position-dependent damping/gain as $\delta=1$ and $\delta=2$, along with the presence of other effects, in two different subsections.
\subsection{Case I: Order-1 $\mathcal{PT}$-symmetric Li\'enard system ($\delta=1$)}
\begin{figure}
	\centering
	\includegraphics[width=0.85\columnwidth]{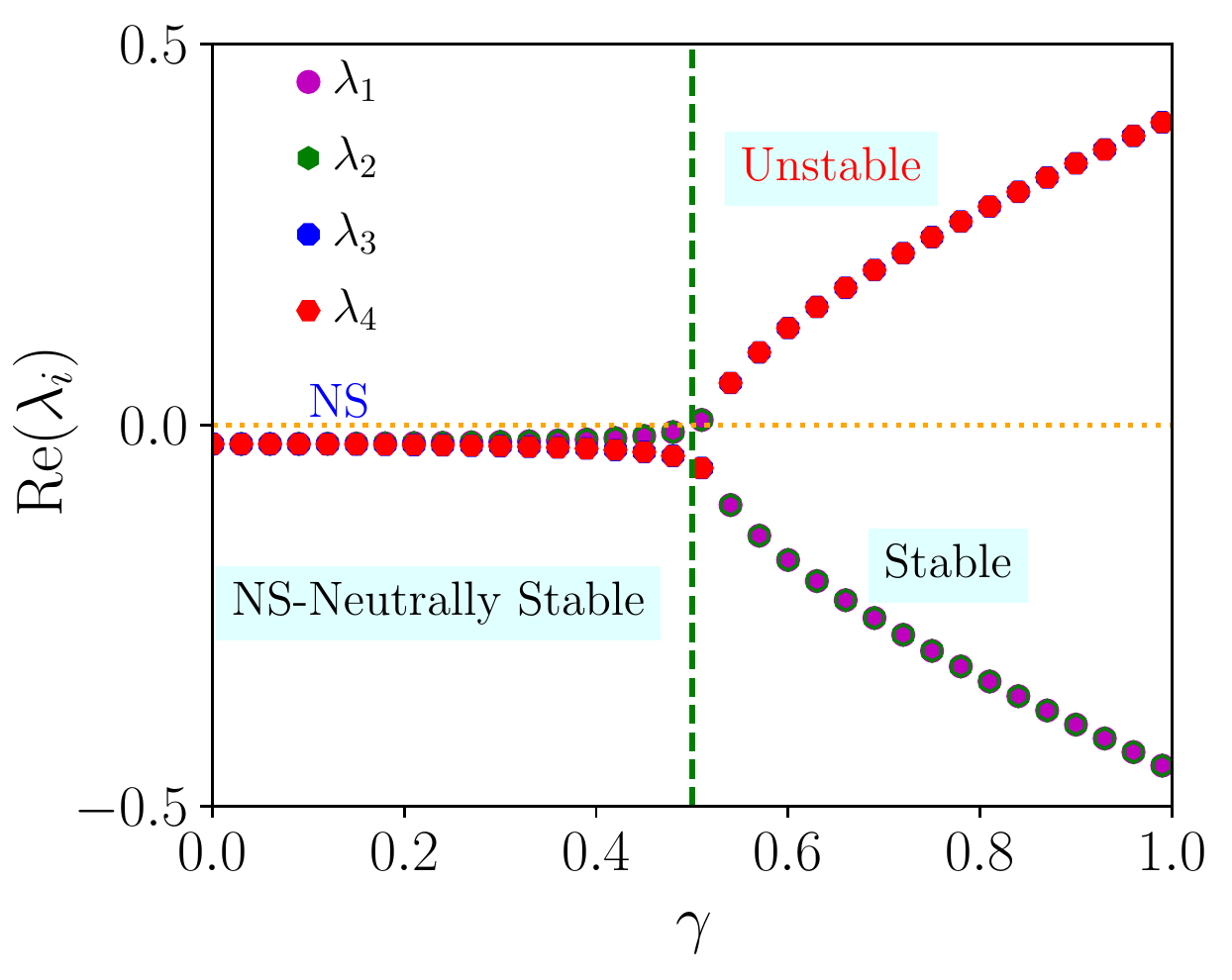}\\
	\includegraphics[width=0.85\columnwidth]{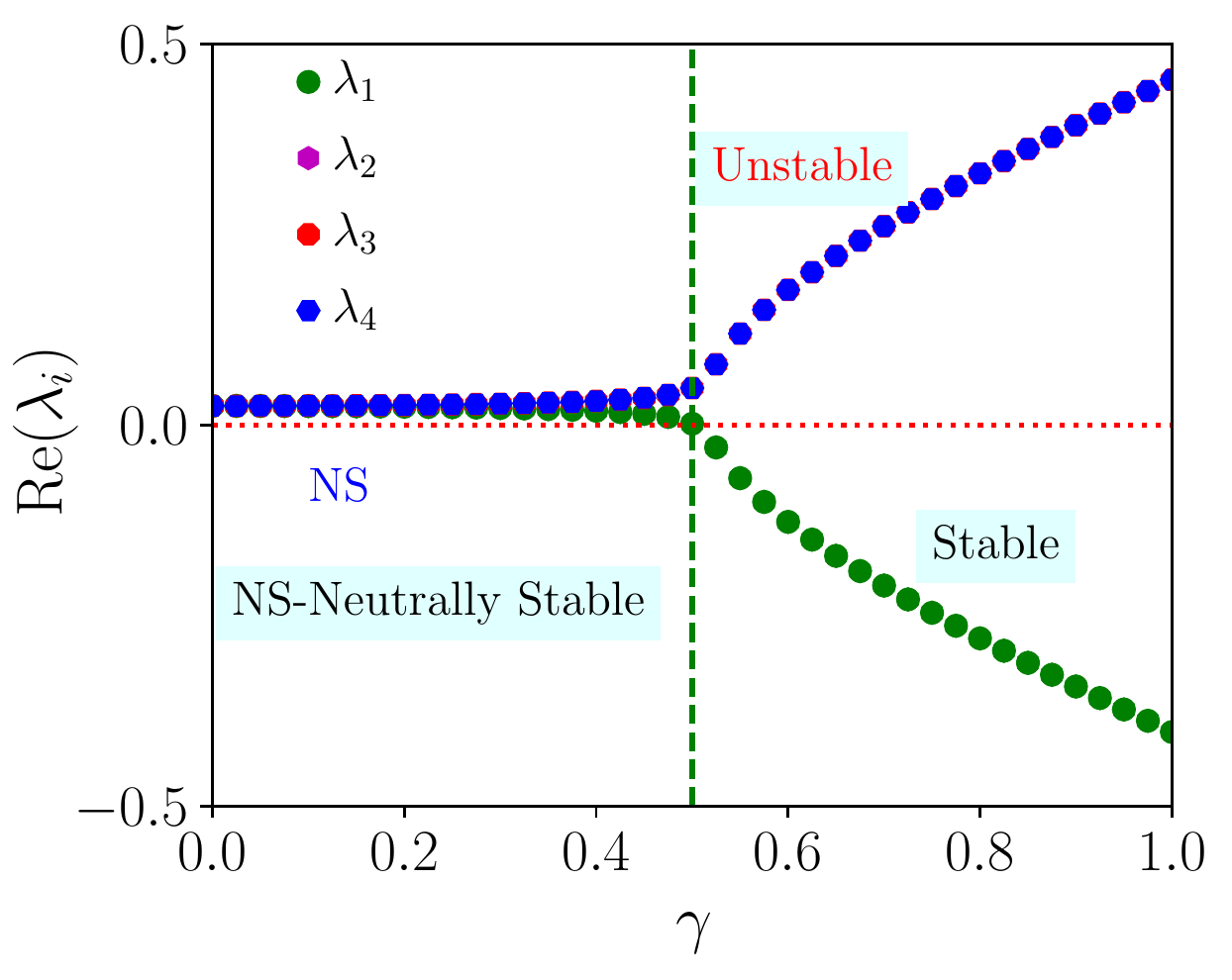}
	\caption{Real component of the eigenvalues of Jacobian for $\mathcal{PT}$-symmetric Li\'enard oscillator.  The plots are drawn for (a) \textbf{FP1} and (b) \textbf{FP2}. Other parameters are assigned as, $\delta=1$ and $\omega_0^2=0.25$.}
	\label{fig1}
\end{figure}
We first analyze the real part of the eigenvalues of the Jacobian  given in Eq. \eqref{eq5} for the order -1 model ($\delta=1$). It must be recalled that the spontaneous symmetry breaking is a typical phenomenon of any $\mathcal{PT}$-symmetric systems where the eigenvalues remain in the neutrally stable region for particular values of some parameters, especially, as a variation of gain/loss and when the gain and loss is increased further, the symmetry breaking takes place by manifesting into asymmetric states featuring two opposite stabilities. Such a dynamics of the proposed system is shown in Fig. \ref{fig1} under the variation of linear loss and gain ($\gamma$) for the two equilibrium points  \textbf{FP1} and \textbf{FP2}. In quite contrast to the conventional symmetry breaking, the plot (see Fig. \ref{fig1}(a)) drawn for the real part of eigenvalues of the Jacobian unravels a different bifurcation where one can observe that the stationary states of the system do not posses any neutrally stable regions for the low values of gain and loss but exhibit a stable state for all the eigenvalues until $\gamma\le0.5$. An opposite case is noticed for the second equilibrium state (\textbf{FP2}) as seen in Fig. \ref{fig1}(b), where the system remains in the unstable state completely for the same values of linear gain and loss. However, in both Figs. \ref{fig1}(a) and \ref{fig1}(b), after this critical value ($\gamma>0.5$), the stability of the system transits into the well-known asymmetric states with a pair of  eigenvalues become stable (as the real components have negative eigenvalues, $\Re(\lambda)<0$) while the other stays in the unstable states as $\Re(\lambda)>0$. Such an interesting observation is a new outcome in the context of $\mathcal{PT}$-symmetric systems.

We now study the temporal dynamics of $\mathcal{PT}$-symmetric oscillators when the gain oscillator is initially excited alone.
\begin{figure} [t]
	\centering
	\includegraphics[width=1\columnwidth]{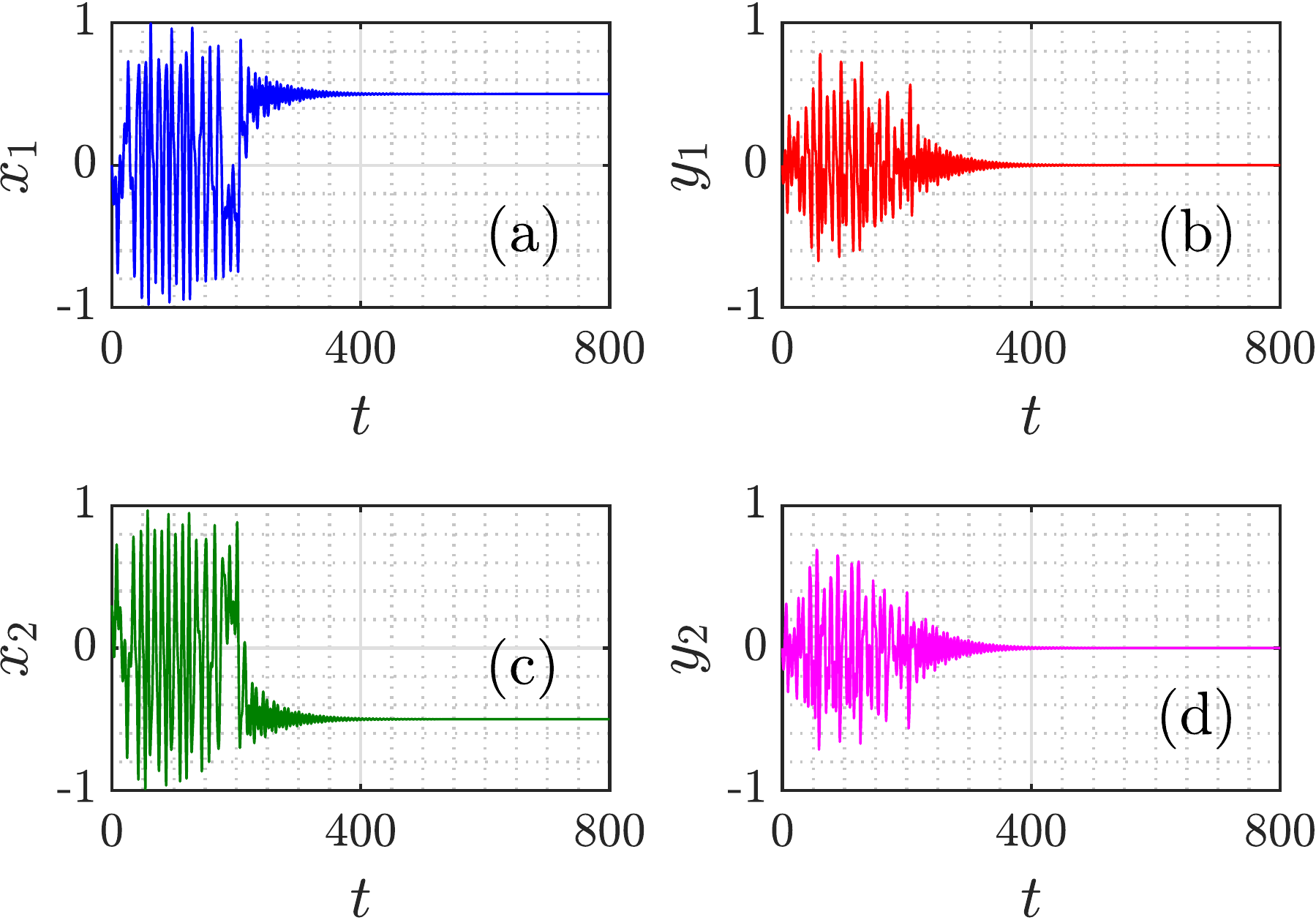}
	\caption{Temporal evolution of $\mathcal{PT}$-symmetric oscillators for $\delta=1$ and $\gamma=0.4$. Top panels (a, b) refer to the time series of lossy oscillators while the bottom panels (c, d) are drawn for the gain oscillators. In both panels, the plots depict the simultaneous occurrence of amplitude and oscillation deaths.}
	\label{fig2}
\end{figure}
To do so, Eq. \eqref{eq4} is numerically integrated by the standard Runge-Kutta fourth order method by considering the linear gain/loss coefficient  and the natural frequency as $\gamma=0.4$, $\omega_0^2=0.25$. Note that when $\gamma=0.4$ (see Fig. \ref{fig1}(a)), the real components of the eigenvalues for \textbf{FP1} are $Re(\lambda_1)=Re(\lambda_2)=-0.018$ and $Re(\lambda_3)=Re(\lambda_4)=-0.031$ and these values corresponding to the second equilibrium point \textbf{FP2} can be read as, $Re(\lambda_1)=Re(\lambda_2)=0.018$ and $Re(\lambda_3)=Re(\lambda_4)=0.031$ as observed in Fig. \ref{fig1}(b). Hence, when the system is given an initial excitation to the gain oscillator as, $\left(x_1(0)=y_1(0)=0, x_2(0)=0.1,y_2(0)=0\right)$, one can observe that the temporal evolution of both gain and lossy oscillators instantaneously approach the stable fixed point (\textbf{FP1}) by manifesting quenching phenomena, as seen in Fig. \ref{fig2}. In particular, it is important to note that in the absence of coupling, the stationary state $\textbf{FP1}$ is a purely imaginary quantity and the only stationary state that exists then in the trivial stationary state $(x_1,y_1,x_2,y_2)=(0,0,0,0)$. Furthermore, it has been mentioned that one of the oscillator is an amplifying system and the other is a lossy system and that the initial excitation is applied to the amplifying oscillator. In such a scenario, there will be blow-up dynamics in the gain oscillator. But due to the presence of coupling between the two oscillators, a stable stationary state exists and quenching phenomenon is observed in the two oscillators. Hence, it could be said that in the present $\mathcal{PT}$-symmetric oscillator, these unique temporal evolutions happen simultaneously in lossy and gain oscillators, which can be regarded as one of the exotic phenomena of $\mathcal{PT}$-symmetric oscillators. Also, it is worthwhile to mention that quenching phenomena such as oscillation and amplitude death usually occur in nonlinear dynamical systems as a consequence of the delay in coupling or some changes in the parameters of the system \cite{r36,r37,r38,r39}. But, in our system, these temporal dynamics emerge as a consequence of parity and time symmetry, which requires the coupling of two oscillator with balanced loss and gain.

\begin{figure}
	\centering
	\includegraphics[width=0.95\columnwidth]{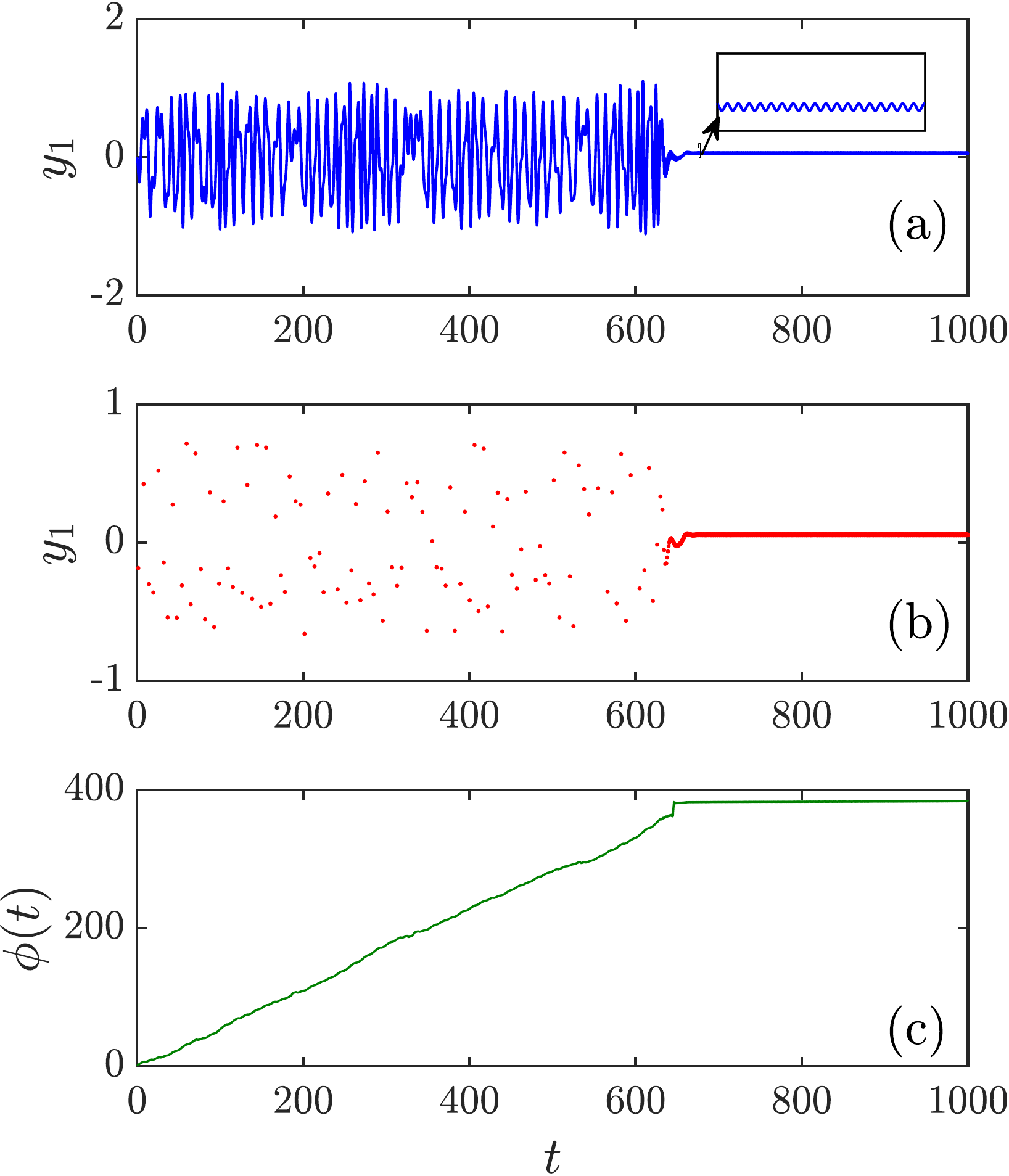}
	\includegraphics[width=1.1\columnwidth]{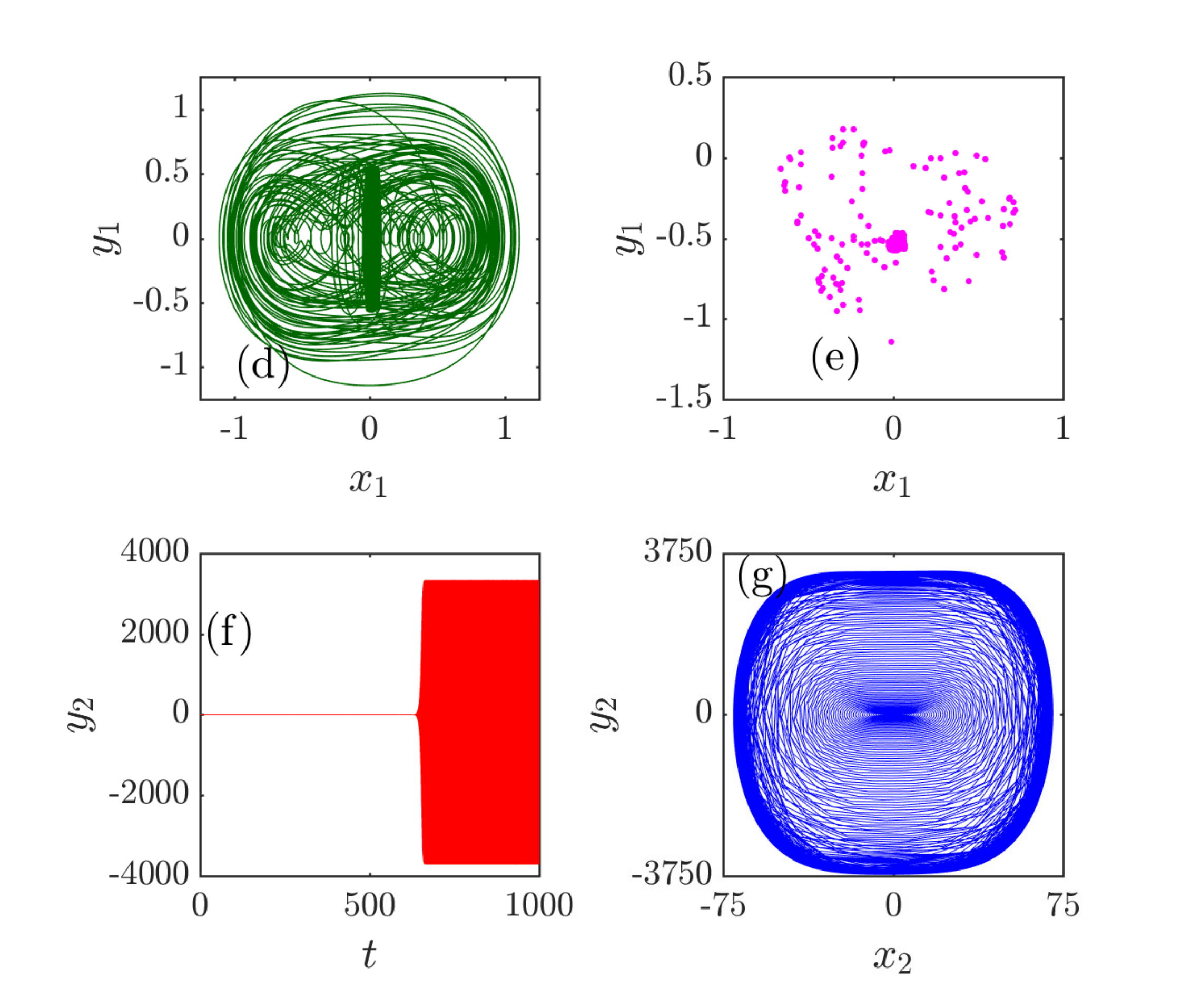}
	\caption{Occurence of transient chaos in lossy oscillators (a-e) and blow-up dynamics in gain oscillators (f-g) for the parameters $\gamma=0.4$, $\omega_0^2=0.25$ and $\delta=1$. Time series showing the transient chaos (a), its corresponding Poincar\'{e} sections (b) and phase of state variable using Hilbert transform (c) are shown. Their phase portraits in ($x_1,y_1$) plane and pertinent Poincar\'{e} maps are depicted in (d) and (e), respectively. The bottom panels (f, g) delineate the phenomenon of blow-up dynamics observed in the gain oscillators.}
	\label{fig4}
\end{figure}

Nevertheless, the dynamics is completely different when the system parameters or initial conditions are changed.  For instance, when the initial condition has been increased further from 0.1 to 0.5 in the gain oscillator as $\left(x_1(0)=y_1(0)=0, x_2(0)=0.5,y_2(0)=0\right)$, we observe the emergence of transient chaos \cite{r39a,r39b,r39c,r39d,r39e} and blow-up dynamics in the loss and gain oscillators, respectively, as shown in Figs. \ref{fig4}(a) and \ref{fig4}(e). It is obvious that in transient chaotic dynamics, the time series undergoes a transition in its behavior from chaotic to periodic as the system is allowed to evolve. 
The transient chaos occurs first, until about $t=640$, when the lossy oscillator settles down at low-amplitude regular oscillation and the gain oscillator blows up. In passing, it is worthwhile to note that Grebogi, Ott and Yorke’s discovery of the emergence of transient chaotic dynamics in systems passing through crisis led to the birth of this field in nonlinear dynamics \cite{r39c}. In particular, to characterize the transient chaos further, we have also shown the Poincar\'{e} maps in ($x_1, y_1$) plane and phase slip of the time series using Hilbert transform, which are, respectively, portrayed in Figs. \ref{fig4}(b) and \ref{fig4}(c). As seen in Fig. \ref{fig4}(b), the chaotic dynamics has been corroborated by the fractal like behaviour wherein one can notice a completely disconnected set of points pertaining to the geometric profile of the time series while the periodic evolutions have been confirmed by the densely interwoven data points found at time series of $t>640$. Also, the phase ($\phi(t)$) of time series can be calculated by using Hilbert transformation which can be defined for a given real valued function g(t),

\begin{equation}
\hat{g}(t)=\frac{1}{\pi}P\int_{-\infty}^{+\infty}\frac{g(\tau)}{t-\tau}d\tau 
\end{equation}
where $P$ is the Cauchy principal value. After applying the Hilbert transform, one can then measure the phase unwrapped by approximately $2\pi$ jumps in the time evolution of the function $\hat{g}(t)$. Figure \ref{fig4}(c) depicts such a calculated phase of time series for the corresponding dynamics of transient chaos where we observe the exponential increase of the phase $\phi(t)$ as the time increases and saturates at a particular time  ($t>640$) and retains in the same phase. These observations, particularly the linear increase indicates the chaotic nature whereas the saturation and remaining in the same phase clearly confirms the periodic evolution. Such a transient chaos can also be verified with the phase portraits drawn in ($x_1,y_1$) plane and their corresponding Poincar\'{e} sections shown in Figs. \ref{fig4}(d) and \ref{fig4}(e), respectively. On the other hand, the gain oscillator exhibits the blow-up dynamics by exponentially increasing to extremely high values (see Figs. \ref{fig4}(f) and \ref{fig4}(g)). This could be attributed to the fact that the fixed point \textbf{FP1} is not a globally stable fixed point. Moreover, it can be found that the initial conditions correspond to the excitation of the gain oscillator. Thus, if the initial excitation is imparted to the gain oscillator, whose value is far away from the stable attractor one, we observe the occurrence of blow-up dynamics in our $\mathcal{PT}$-symmetric coupled systems.

\begin{figure}
\centering
\includegraphics[width=1\columnwidth]{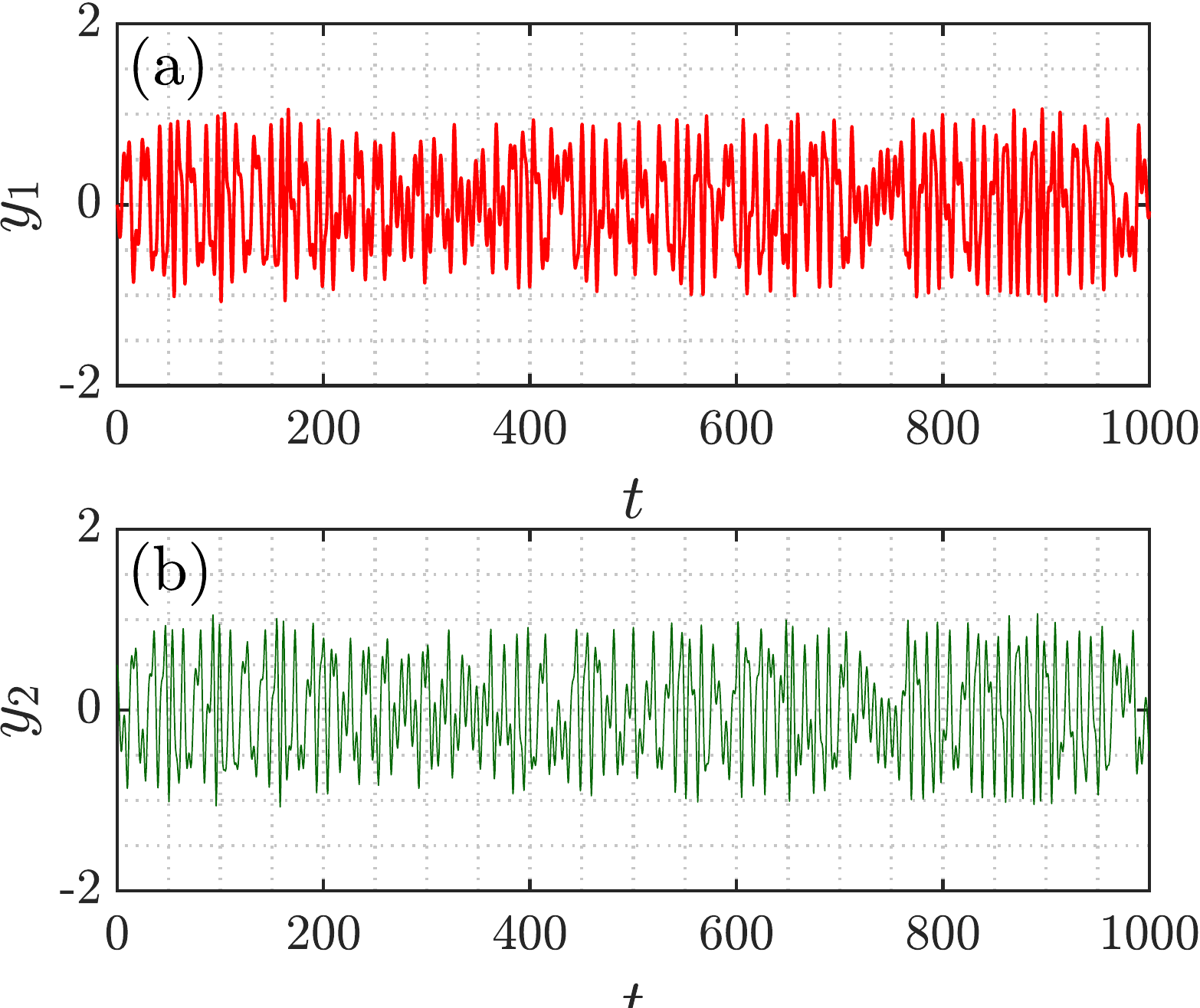}\\
\includegraphics[width=1\columnwidth]{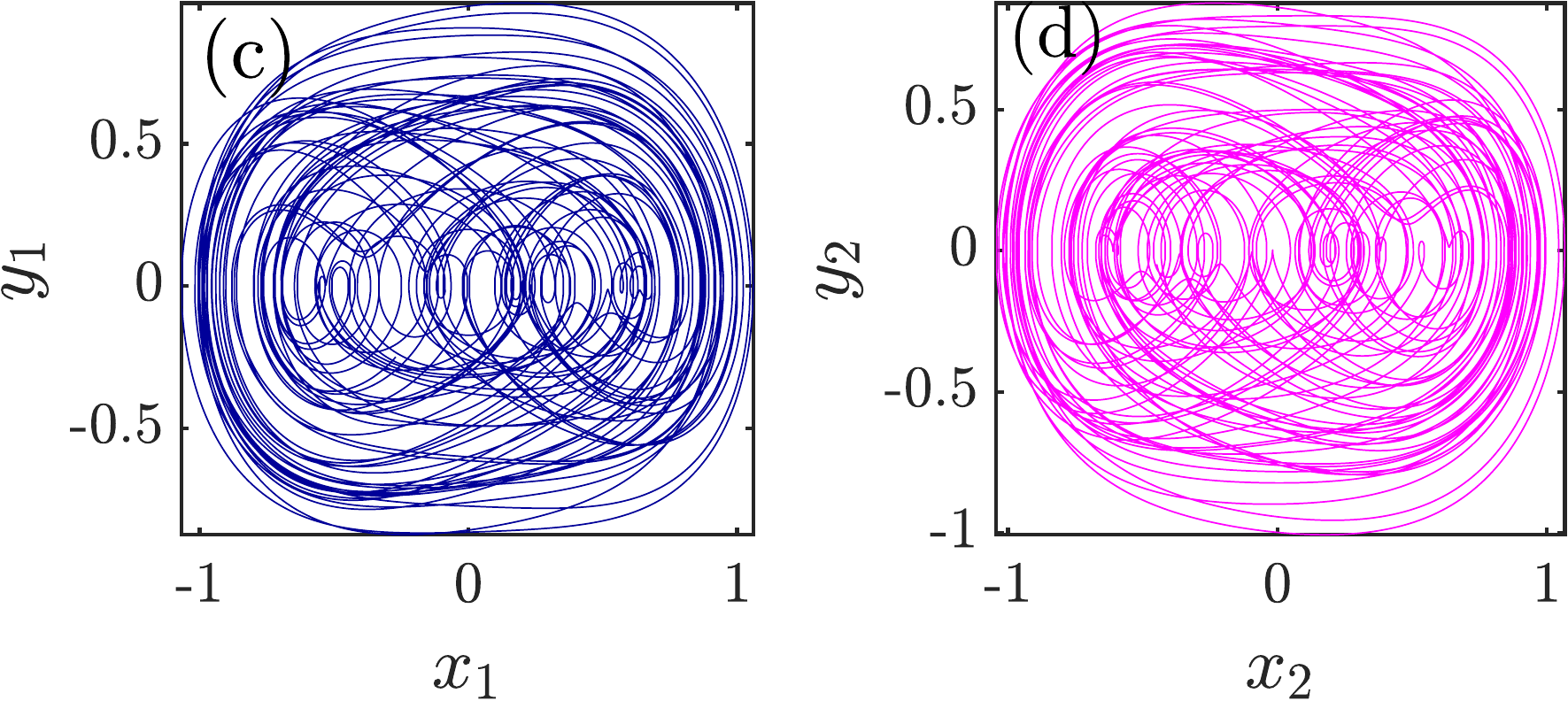}
\caption{Temporal evolution of $\mathcal{PT}$-symmetric oscillators for (a) lossy and (b) gain oscillators when the parameters are assigned as $\gamma=0.4$, $\delta=1$ and $\omega_0^2=0.25$. The external source parameters are: $A_0=0.0001$, $\omega=2\pi/500$ and $\psi=10^{-5}$. The plots (c, d) show the corresponding phase trajectories in ($x_i,y_i$, $i=1,2$) planes.}
\label{fig5}
\end{figure}

It is to be remembered that the blow-up dynamics is a typical outcome of most of the $\mathcal{PT}$-symmetric systems, especially when the system is set to be operated above the $\mathcal{PT}$-symmetric threshold (broken $\mathcal{PT}$-symmetry) and one has to seek how the blow-up dynamics can be controlled in order to make the system more feasible \cite{r6}. We now present some possible ways to tame the blow-up dynamics in such $\mathcal{PT}$-symmetric oscillators. In particular, we show how such a blow-up dynamics in the $\mathcal{PT}$-symmetric Li\'enard system can be controlled using an external driving force. We consider the external force to be of the form $f(t)=f_0 \cos(\omega t + \psi t^3)$, where $f_0$ is the amplitude, $\omega$ is the angular frequency of the driving force and a strength of chirped modulation of the external drive is noted through a parameter $\psi$. As seen in Fig. \ref{fig5}, it is straightforward to notice that when the lossy oscillator is exited using the chirped cosine drive, the blow-up dynamics could be removed in the gain oscillator and the former transits to aperiodic evolution.  Similarly, the transient chaos observed in lossy oscillators (cf. Fig. \ref{fig4}(a)) has also been transformed into a complete chaotic dynamics. From our further investigations, we find that controlling blow-up dynamics is possible only via modulation of the strength of the external chirp.
It is worthwhile to mention that such modulation of external chirp is recently reported to be utilized in controlling chaos in ultra-cold systems \cite{r40}. Hence, it could be ascertained that the blow-up dynamics in $\mathcal{PT}$-symmetric Li\'enard type oscillators can be completely arrested by externally driving the system.

\begin{figure}
\centering
\includegraphics[width=0.8\columnwidth]{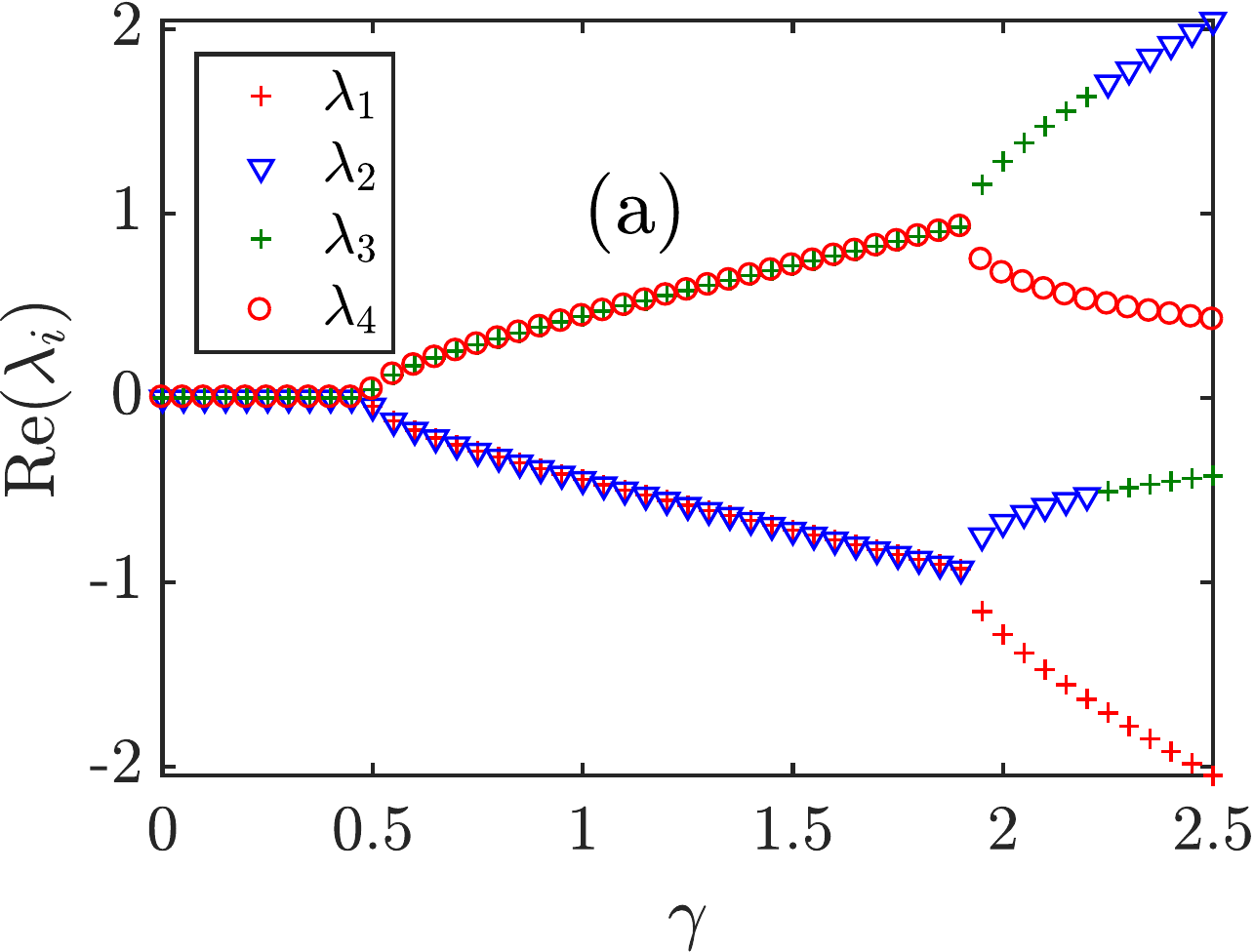}\\
\includegraphics[width=0.8\columnwidth]{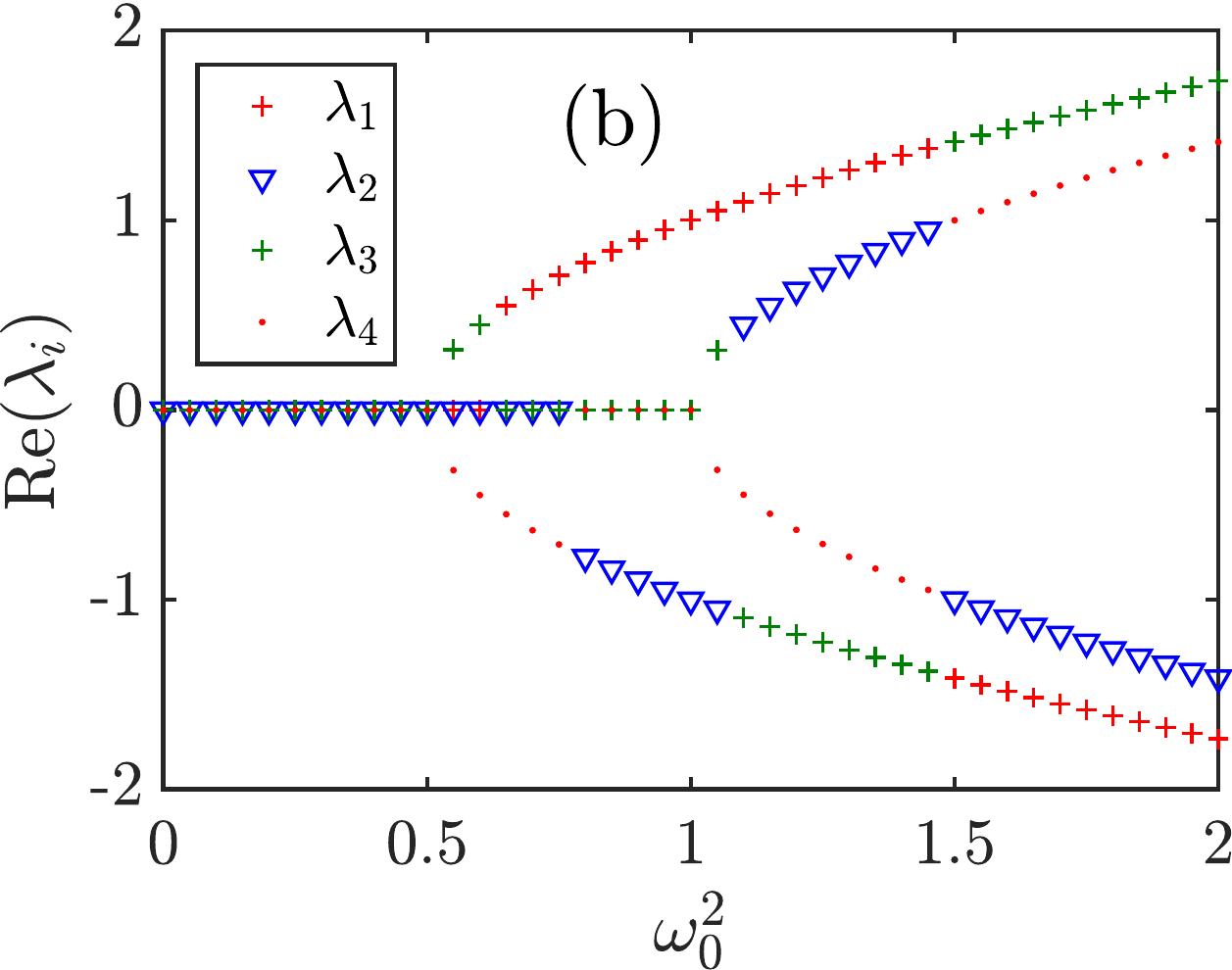}\\
\caption{Real component of the eigenvalues of Jacobian for $\mathcal{PT}$-symmetric Li\'enard oscillator. The plots are drawn as a function of (a) linear gain/loss ($\gamma$) and (b) natural frequency ($\omega_0^2$) for \textbf{FP1}. Other parameters: (a) $\omega_0^2=0.25$ and (b) $\gamma=0.4$. In both figures, $\delta=2$.}
\label{fig6}
\end{figure}

\begin{figure}[t]
	\centering
	\includegraphics[width=1\linewidth]{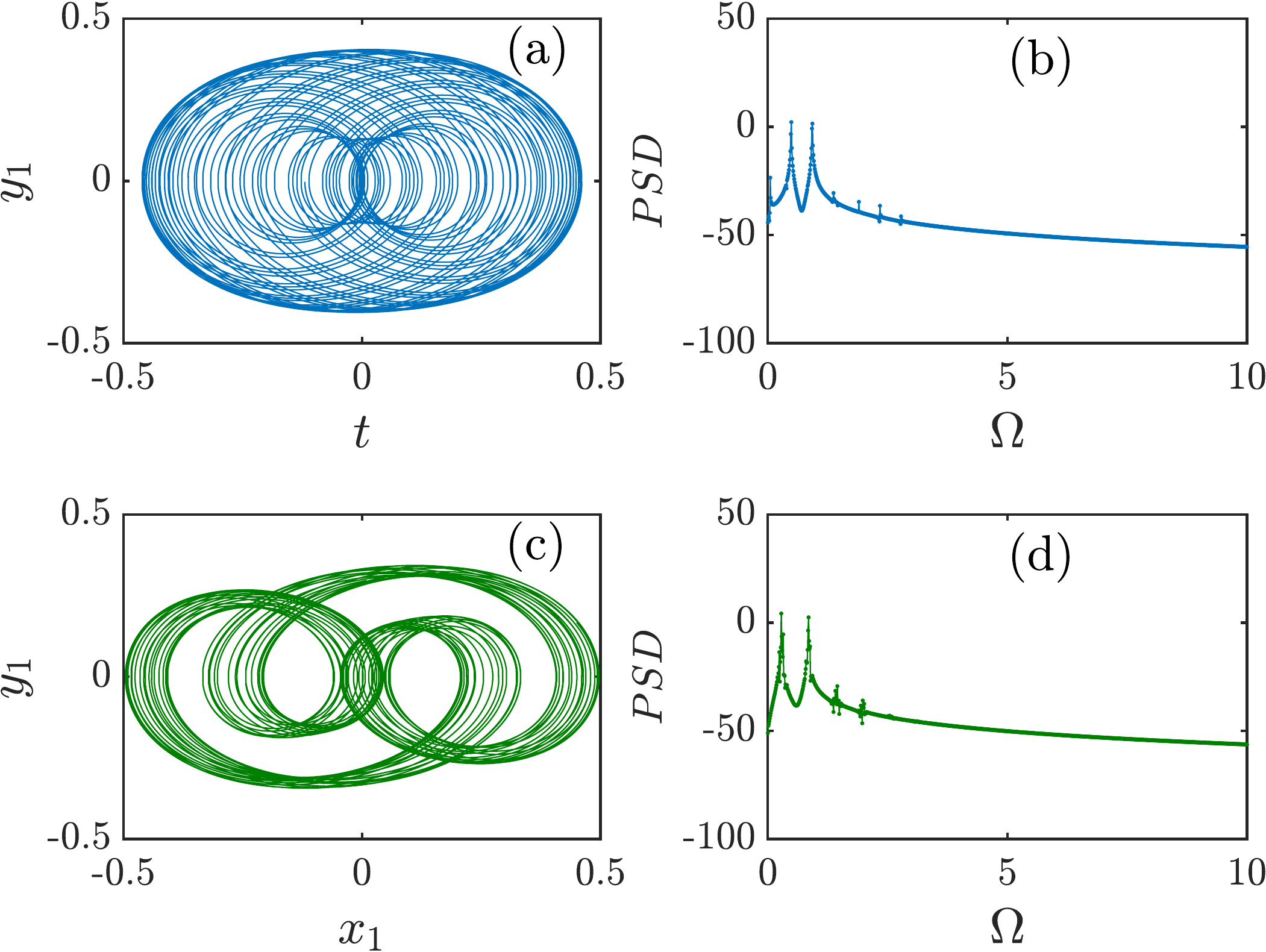}
	\caption{Phase portraits, drawn in ($x_1,y_1$) plane, showing the quasiperiodic behaviour of the $\mathcal{PT}$-symmetric Li\'{e}nard systems (gain oscillators) when (a) $\omega_0^2=0.75$  and (c) $\omega_0^2=0.5$. Right panels depict the corresponding power spectral density. Other parameters: $\gamma=0.1$ and $\delta=2$.}
	\label{fig7}
\end{figure}

\subsection{Case II: Higher order (nonlinear damping) $\mathcal{PT}$-symmetric Li\'enard system ($\delta=2$)}

For the higher order system too, we start with the study of stability analysis of the eigenvalues of Jacobian associated with the equilibrium points \textbf{FP1} as a function of linear gain/loss coefficient $\gamma$ and natural frequency $\omega^2_0$. Unlike the order-1 model, the higher model exhibits the typical symmetry breaking bifurcations as shown in Fig. \ref{fig6}.
As can be seen in Fig. \ref{fig6}(a), the eigenvalues of the Jacobian are found to be purely imaginary  (in other words, the real parts of all the eigenvalues become zero, i.e., $\Re(\lambda)=0$) and so the stationary state of the system remains in the neutrally stable regime until the gain and loss attain some non-trivial values ($\gamma\le0.5$). After this critical value, a pair of eigenvalues attain positive values whereas other pair become negative, which, respectively, transforms the stationary state into a saddle fixed point. However, an interesting feature is noticed when the gain and the loss parameter is fixed at $\gamma=2$, where two of the eigenvalues once again bifurcate with a swapping of eigenvalues as they take positive and negative values abruptly.
On the other hand, the eigenvalues of the Jacobian as a function of $\omega^2_0$ also depict the same characteristics with the stability threshold situated at $\omega^2_0=0.5$ (see Fig. \ref{fig6}(b)). Nevertheless, one can observe that these eigenvalues swap between positive and negative real components once they reach the stability threshold. In particular, once crossed this critical value, a pair of eigenvalues remain in neutrally stable regime by exhibiting a pure imaginary eigenvalues ($\Re(\lambda)=0$). While other pair of real parts of eigenvalues start to bifurcate by transforming into a saddle one (having one positive and one negative real component). Such a novel phase transition of eigenvalues is completely unusual in the context of nonlinear dynamical systems.

\begin{figure*}[t]
	\centering
	\includegraphics[width=1.7\columnwidth]{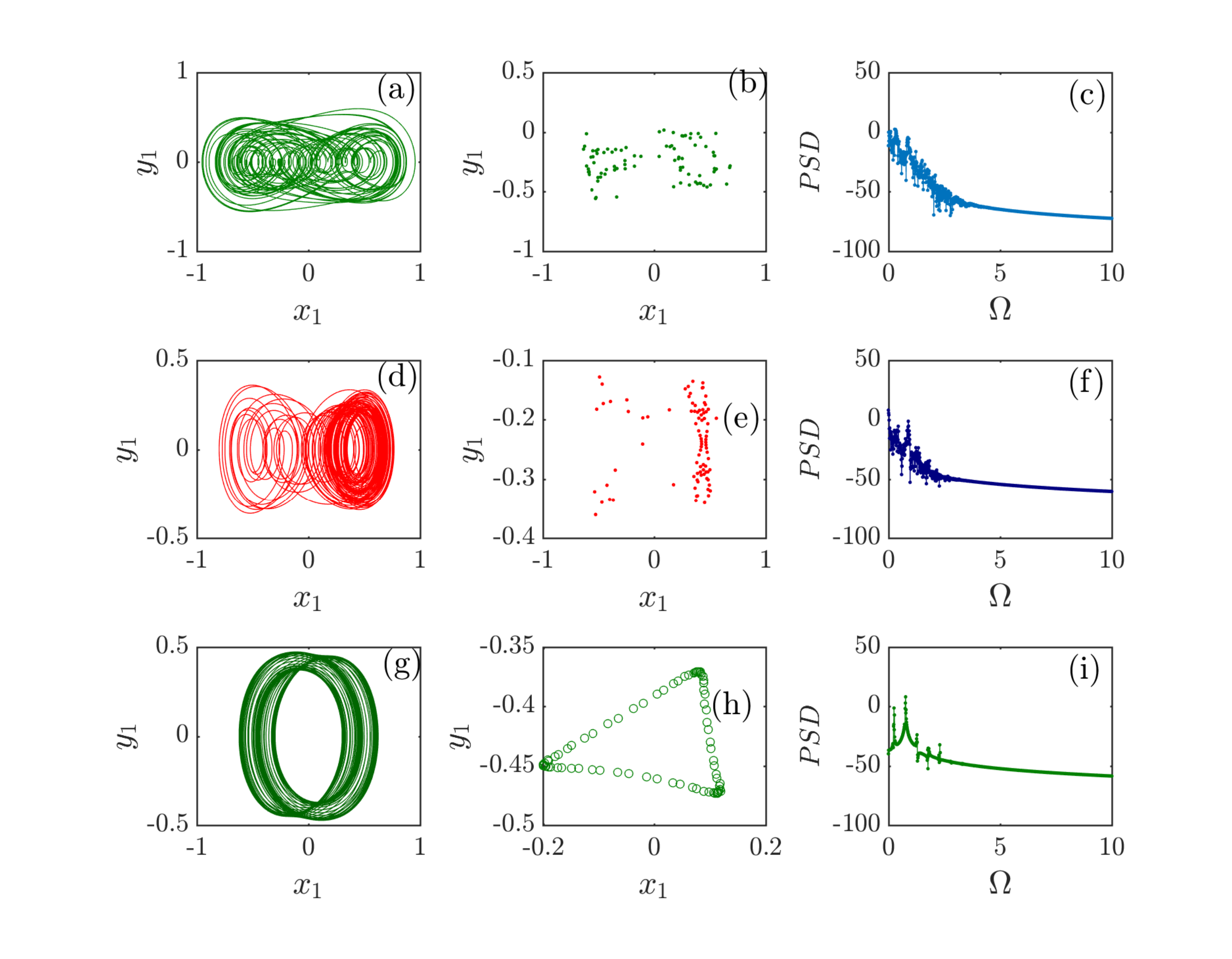}\\
	\includegraphics[width=1\columnwidth]{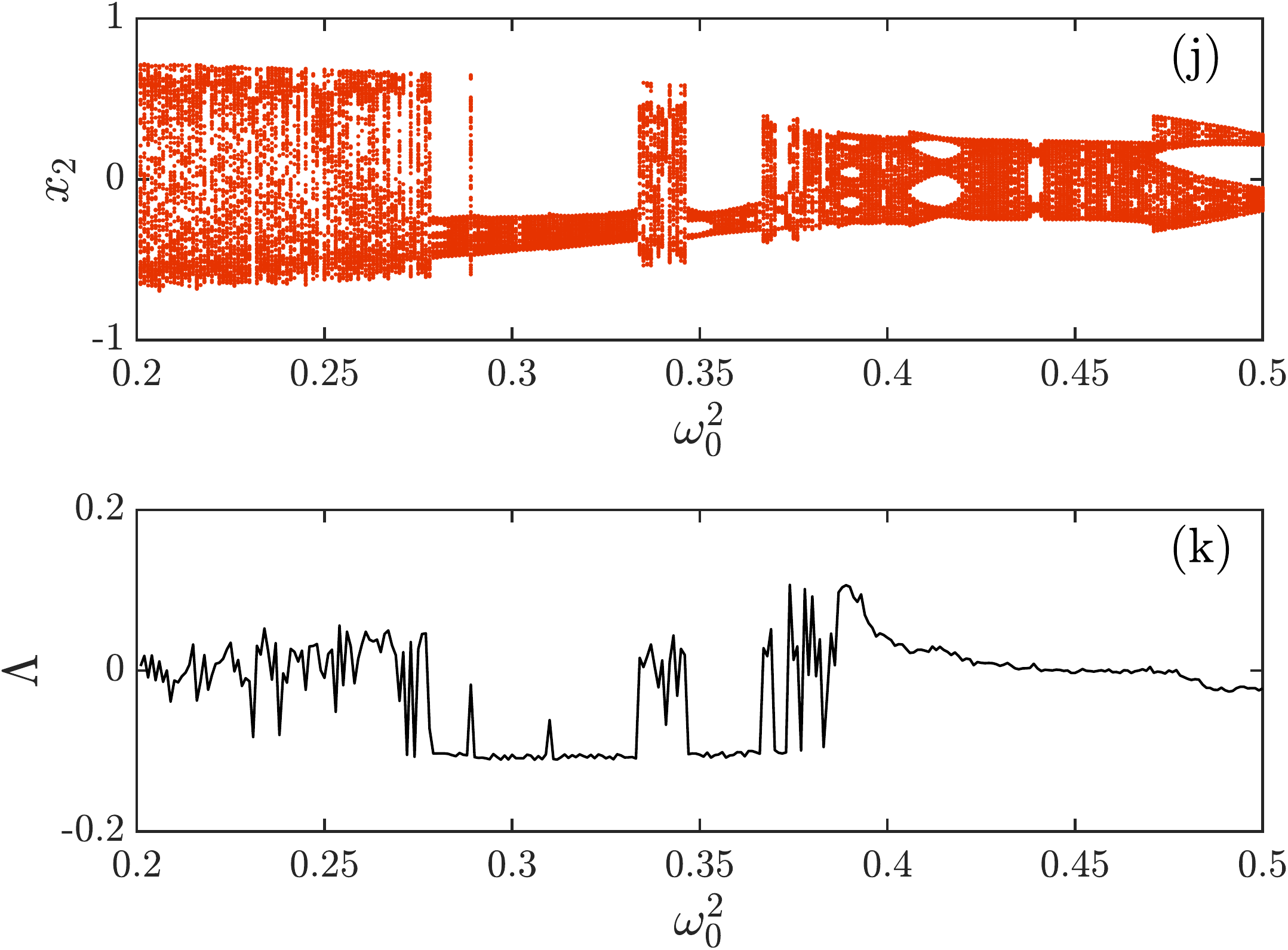}
	\caption{Chaotic dynamics as a function of initial conditions and control parameters are shown for (a) when the gain oscillator alone is initial exited, (d) lossy oscillator is excited and (f) both gain and lossy oscillators are simultaneously exited. The plots (b, e and h) show the Poincar\'{e} maps in ($x_1,y_1$) plane whereas the plots (c, f and i) indicate the corresponding power spectra.  The two bottom panels (j, k) portray the one parameter bifurcation diagram  and maximal Lyapunov exponent for the gain oscillators as a function of natural frequency. The other parameters are $\omega_0^2=0.25$, $\gamma=0.1$ and $\delta=2$.}
	\label{fig8}
\end{figure*}

We would now discuss the temporal evolution of the higher order $\mathcal{PT}$-symmetric oscillator for three values of natural frequency ($\omega_0^2$). Note that order-1 system exhibit the rich nonlinear dynamics including quenching phenomena, transient chaos and the blow-up dynamics (which is a typical outcome of any $\mathcal{PT}$-symmetric ordinary differential equation systems \cite{r33}). We now investigate the temporal dynamics of the higher order nonlinear damping model.To characterize chaotic dynamics in the system, we make use of the power spectral analysis method and this could be calculated from the Fourier series of the time series as given below \cite{r41}:
\begin{equation}
P(\Omega)=|\tilde{x}_j(\Omega)|^2, \quad\mathrm{where}, \quad j=1, 2.
\label{eq7}
\end{equation}
In Fig. \ref{fig7}, we have plotted the phase plane trajectory of the gain oscillator (the lossy oscillator also exhibits the same mirror dynamics and they were not shown here and in forthcoming figures) and its corresponding power spectra for $\gamma=0.1$ and $\omega^2_0=0.75$ (which lies in the saddle region, see Fig. \ref{fig6}(b)). The power spectra (Fig. \ref{fig7}(b)) show two distinct sharp peaks at $\Omega_1=0.49$ and $\Omega_2=0.93$ and their ratio becomes $\Omega_{2}/\Omega_{1}\approx1.8979$, meaning that the temporal dynamics of the gain oscillator exhibits the quasiperiodic behaviour. Moreover, from the phase space trajectory (Fig. \ref{fig7} (a)), one can notice that it features a 2-torus behaviour. The dynamics gets altered if we change the value of natural frequency which is shown in Fig. \ref{fig7}(d),  where we have plotted the power spectra for $\omega^2_0=0.5$. The power spectra once again delineate two distinct sharp peaks at $\Omega_1=0.28$ and $\Omega_2=0.85$. and their ratio is $\Omega_2/\Omega_1\approx3.0357$, which is once again incommensurate. With this observation, we show that on decreasing the value of $\omega^2_0$, the frequency ratio could be held incommensurate. Further, as $\omega^2_0$ is decreased further to $\omega^2_0=0.25$, we find that the whole low frequency region is populated and there is an exponential decay in the higher frequency region (see Fig. \ref{fig8}(c)). This dynamics certainly depicts the chaotic behavior of our proposed $\mathcal{PT}$-symmetric system and hence we can conclude that the system exhibits the typical quasiperiodic route to chaos \cite{r41}. Moreover, it must be noted here that the transition in the phase plane trajectory from a quasiperiodic to chaotic dynamics occurs as a result of the decreasing the value of $\omega^2_0$ from the saddle node to the neutrally stable regime. In other words, it could be inferred that the emergence of a quasiperiodic toroidal attractor results as a result of symmetry breaking. 

To analyze the temporal dynamics in detail for $\omega_0^2=0.25$ and different initial conditions, the phase plane trajectory of the chaotic attractor has been plotted in Fig. \ref{fig8} for three cases, \textit{viz.} when the gain oscillator is excited alone, when the loss oscillator is excited alone and when both oscillators are simultaneously excited. For initial launch conditions conforming to the first case ($\left(x_1(0)=y_1(0)=0, x_2(0)=0.45,y_2(0)=0\right)$), one can observe that the chaotic dynamics resembles like a Lorenz-type chaotic attractor. The dynamical scenario is completely different if the lossy oscillator has been excited ($\left(x_1(0)=0.45, y_1(0)=0, x_2(0)=y_2(0)=0\right)$) instead of the gain counterpart where one observes that the double band chaos is occurred with one chaotic band is being populated densely which mainly serves as an attractor while other band acts a repeller. If both gain and loss oscillators are excited simultaneously with  same initial conditions $\left(x_1(0)=0.45, y_1(0)=0, x_2(0)=0.45, y_2(0)=0\right)$, the dynamics of the system transforms into quasiperiodic one, as shown in Fig. \ref{fig8}(g). We have also presented the one parameter bifurcation and corresponding maximal Lyapunov exponent for the variation of the natural frequency parameter ($\omega_0^2$) when the gain oscillator is given an initial excitation. From the figure, it could be clearly seen that on decreasing the value of the natural frequency, the $\mathcal{PT}$-symmetric coupled oscillators show a much complicated dynamics, revealing the coexistence of quasiperiodic and chaotic dynamics. Especially, one can observe that the transition takes place from quasi periodic windows to periodic orbits and to chaotic ones. With these observations,  it is very important to stress that the higher order $\mathcal{PT}$-symmetric systems transit the dynamics from quasiperiodic to chaotic one depending on the nature of the control system parameters or the initial conditions. 
\section{EXPERIMENTAL REALIZATION}
\label{sec4}
\begin{figure}[t]
\centering
\includegraphics[width=1\linewidth]{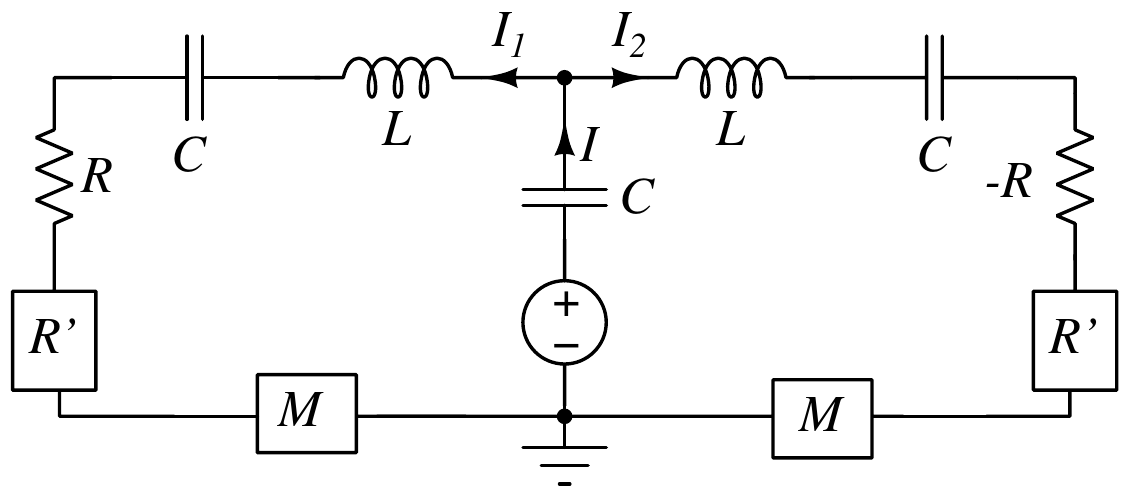}
\caption{Schematic representation of the electronic circuit. Here, $L$, $C$ and $R$ are the inductor, capacitor and the resistor, respectively. $R^{'}$ is th non-ohmic resistor and $M$ is the current multiplier.}
\label{fig9}
\end{figure}

In this section, we provide a possible experimental scheme to realize the proposed system studied in the previous sections.
Using Kirchoff’s law, and in the absence of any resistive element and current multiplier, the equations governing the dynamics in electric current in a CLC circuit (see schematic Fig. \ref{fig9}) are given by
\begin{subequations}
\begin{align}
L \frac{d I_1}{d t} + \frac{Q_1}{C} + \frac{Q}{C} = V(t) \\
L \frac{d I_2}{d t} + \frac{Q_2}{C} + \frac{Q}{C} = V(t)
\end{align}
\label{eq6}
\end{subequations}
where $Q$ , $Q_1$  and $Q_2$  are charges in the capacitors. From the circuit diagram (\ref{fig9}), we can note that the electric charge in the capacitors can be written in terms of the current flowing in the circuit as follows.
\begin{equation}
\frac{d Q_1}{d t} = I_1, \quad 
\frac{d Q_2}{d t} = I_2, \quad 
\frac{d Q}{d t} = I  
\label{eq7}     
\end{equation}
Hence, we have $Q=Q_1+Q_2$ and using this relation, we can eliminate $Q$ in Eqs. \ref{eq6}(a) and \ref{eq6}(b) and hence they can be rewritten as
\begin{subequations}
\begin{align}
L \frac{d I_1}{d t} + 2 \frac{Q_1}{C} + \frac{Q_2}{C} = V(t) \\
L \frac{d I_2}{d t} + \frac{Q_2}{C} + 2 \frac{Q_2}{C} = V(t)
\end{align}
\label{eq8}
\end{subequations}
Differentiating Eqs. \ref{eq8}(a) and Eq. \ref{eq8}(b) and using Eq. \ref{eq7}, we obtain the following equations
\begin{subequations}
\begin{align}
L \frac{d^2 I_1}{d t^2} + \frac{2 I_1 + I_2}{C}  = V^{'}(t) \\
L \frac{d^2 I_2}{d t^2} + \frac{I_1 + 2 I_2}{C}  = V^{'}(t)
\end{align}
\label{eq9}
\end{subequations}
where $V^{'}=dV/dt$. These equations describe the oscillatory behavior of current in a CLC oscillator, but this system is not $\mathcal{PT}$-symmetric. To ensure the presence of $\mathcal{PT}$-symmetry, we need to include a resistor in both loops. One of the resistors must induce attenuation in the electric current and the other needs to amplify in equal proportion. This can be achieved by using a negative resistance in one half of the circuit. Unlike a normal resistance, the I-V characteristics of a negative resistance shows a decrease in the current flowing through it when the potential difference is increased. A negative resistance can be realized in our circuit by using an \textit{Esaki} or \textit{Gunn} diode. Then, from Eqs. \ref{eq9}(a) and \ref{eq9}(b), we have
\begin{subequations}
\begin{align}
\frac{d^2 I_1}{d t^2} + \alpha \frac{d I_1}{d t} + \omega_{0}^2 I_1 + \kappa I_2  = \frac{V^{'}(t)}{L} \\
\frac{d^2 I_2}{d t^2}  - \alpha \frac{d I_2}{dt} +\omega_{0}^2 I_2 + \kappa I_1  = \frac{V^{'}(t)}{L}
\end{align}
\label{eq10}
\end{subequations}
where $\alpha=R/L$, $\omega_0^2=2/LC$ and $\kappa=1/LC$ are the linear gain/loss coefficient, natural frequency of the oscillator and the coupling constant, respectively. 
Until this point, we have not yet included nonlinear resistance and current multiplier in the circuit that would facilitate the realization of our theoretical model. The nonlinear resistance $R_{nl}$ could be realized by using a non-ohmic resistance. In a non-ohmic resistance, the I-V characteristics do not follow a straight line. Instead, the gradient of the graph increases as the current is increased. We choose the nonlinear resistance to follow the generalized I-V characteristics law given by $V_{nl,i}=R_{nl,i}I^\delta$, where $R_{nl}$ is the nonlinear coefficient and $\delta$ is the exponent of nonlinear resistance, respectively. Also, the Duffing nonlinearity ($\chi$) can be easily realized in the circuit by using an analog device multipliers \emph{AD532} or \emph{AD534} \cite{r42}. For current inputs $X$ and $Y$, such a device would produce an output $XY/10$. Then from Eqs. \ref{eq10}(a) and \ref{eq10}(b), we have
\begin{subequations}
\begin{align}
\frac{d^2 I_1}{d t^2} + (\alpha + \beta_1 I_1^{\delta-1})\frac{d I_1}{d t} +\chi I_1^3+ \omega_{0}^2 I_1 + \kappa I_2  = \frac{V^{'}(t)}{L} \\
\frac{d^2 I_2}{d t^2}  - (\alpha + \beta_2 I_2^{\delta-1})\frac{d I_2}{dt} + \chi I_2^3 +\omega_{0}^2 I_2 + \kappa I_1  = \frac{V^{'}(t)}{L}
\end{align}
\label{eq11}
\end{subequations}
where $\beta_i= \delta R_{nl,i} /L$ $(i=1, 2)$ is the coefficient of nonlinear resistance and the term $\chi$ denotes the Duffing nonlinearity. It should be noted that one of the non-ohmic resistors must play the role of a nonlinear amplifier, i.e. $\beta_2=-\beta_1$.\\  
\vspace{-2em}
\section{Conclusion}
\label{sec5}
We have investigated a $\mathcal{PT}$-symmetric Li\'enard oscillator configuration for two cases of nonlinear dissipation. We have analyzed the stability of the stationary states using the Jacobian followed by the linear stability analysis. In the stable region of the first order model, we have found that the choice of initial conditions gives rise to a variety of rich dynamics including amplitude death, oscillation death, transient chaos and blow-up dynamics. Using an external driving force, the blow-up dynamics has been controlled and a pure aperiodic state has also been achieved in the system. 
On the other hand, the second order model exhibits the typical quasi-periodic route to chaos as the natural frequency of the oscillator is decreased from the saddle node regions to stable regime. For the potential experimental realization of the proposed theoretical models, we have presented an electronic circuit using Kirchoff's laws. The gain oscillator can thus be realized using a negative resistance while the nonlinear dissipation can be achieved using a non-ohmic resistance.

We anticipate that these systems could be scaled up to many oscillators, and these novel ramifications will open an interesting avenue to understand the controlling of blow-up dynamics in many body driven-dissipative systems.
\setcounter{secnumdepth}{0}
\section{ACKNOWLEDGEMENTS}
J. P. D. thanks MHRD, Government of India for financial support through a fellowship and A. K. S. acknowledges financial support from Department of Science and Technology (DST) and Science and Engineering Research Board
(SERB), Government of India. A. G. is indebted to DST-SERB, Government of India,
for providing a National Postdoctoral Fellowship (Grant No. PDF/2016/002933). M. K. gratefully acknowledges the Ramanujan Fellowship SB/S2/RJN-114/2016 from the DST-SERB, Government of India. This research was also supported in part by the International Centre for Theoretical Sciences (ICTS) during a visit for participating in the program -- Non--Hermitian Physics --PHHQP XVIII (Code: ICTS/nhp2018/06)
\setcounter{secnumdepth}{0}

\end{document}